\documentclass[runningheads]{llncs}
\usepackage[T1]{fontenc}
\usepackage{graphicx}
\usepackage{listings}
\usepackage{xcolor}
\usepackage{minted}
\usepackage{subcaption}
\usepackage{amsmath}
\usepackage{amsfonts}
\usepackage[T1]{fontenc}
\usepackage[utf8]{inputenc}
\usepackage{cite}
\usepackage{amsmath,amssymb,amsfonts}
\usepackage{textcomp}
\usepackage{booktabs}
\usepackage{array}
\usepackage{hyperref}
\usepackage{url}
\usepackage{tikz}
\usepackage{pgfplots}
\pgfplotsset{compat=1.18}
\usetikzlibrary{calc,shapes.geometric, arrows.meta, positioning, fit, backgrounds}
\newcommand{\ignore}[1]{}

\begin{document}
%
\title{LLMs in DevSecOps: A Systematic Mapping of Models, Evidence, and Research Gaps Across the Software Lifecycle}

%
%
\author{Sindu Chitraju\inst{1}\orcidID{0009-0004-0446-8388} \and
Mir Mehedi A. Pritom\inst{1}\orcidID{0000-0002-1260-1829} 
}
\authorrunning{Sindu et al.}
%
\institute{Department of Computer Science, Tennessee Technological University\\Cookeville, TN 38505, United States\\
\email{\{schitraju42,mpritom\}@tntech.edu}}
\maketitle              
\begin{abstract}

Large language models (LLMs) are increasingly being integrated into software security and DevSecOps workflows, yet their adoption, evaluation, and maturity vary substantially across lifecycle stages. This study systematically maps recent LLM-enabled security research published between 2024 and 2026 (until June) across six DevSecOps stages: \textit{Plan}, \textit{Code}, \textit{Test}, \textit{Review}, \textit{Deploy}, and \textit{Monitor}. For each stage, we analyze the security tasks addressed, models and adaptation strategies employed, evaluation settings, reported effectiveness, and key limitations. Our findings reveal an uneven research landscape, with \textit{Test} and \textit{Code} receiving substantially greater attention than \textit{Deploy} and \textit{Review}. Across stages, recent systems increasingly rely on structured context, iterative reasoning, fine-tuning, retrieval augmentation, or agentic workflows rather than direct single-shot prompting. However, reported performance remains difficult to compare because studies use heterogeneous tasks, datasets, metrics, and baselines. Furthermore, most approaches are evaluated using benchmarks or open-source repositories, while industrial-scale evidence remains limited. We also find that existing systems are predominantly stage-specific, with little evidence of LLM-enabled security reasoning spanning multiple DevSecOps stages. Based on these observations, we identify research opportunities in standardized evaluation, \\production-scale validation, supply-chain and Infrastructure-as-Code security, trustworthy autonomous remediation, and cross-stage LLM integration. This study provides a recent lifecycle-level view of where LLM-enabled DevSecOps research is concentrated, what methodological patterns are emerging, and where significant gaps remain.

\keywords{Large Language Models \and DevOps \and DevSecOps \and SDLC Security \and Software Security \and LLMs in Software Security}

\end{abstract}

\section{Introduction}
\label{sec:intro}

Large language models (LLMs) are increasingly being integrated into software engineering and cybersecurity workflows, enabling automation and decision support across the software development lifecycle (SDLC) \cite{fu2025b,zhang2025llms,zhou2024}. In DevSecOps, security activities span the full lifecycle: \textit{Plan}, \textit{Code}, \textit{Test}, \textit{Review}, \textit{Deploy}, and \textit{Monitor}, including threat modeling, vulnerability detection and repair, security testing, code review, CI/CD management, and runtime monitoring \cite{fu2025b}. Recent studies have applied LLMs to many of these tasks, including threat modeling \cite{bedoya2024,yang2024}, vulnerability repair \cite{fakih2025,kim2025,nong2025}, secure code review \cite{liu2026securereviewer,yu2026}, CI/CD diagnosis \cite{xu2025logsage}, and anomaly detection and remediation \cite{guan2025,jha2025,kakarla2024}.

Despite this rapid adoption, the current research landscape remains fragmented. Existing studies differ substantially in the tasks they address, models and adaptation strategies they use, datasets and metrics they evaluate on, and the extent of real-world validation they provide \cite{ding2024,fu2025b,zhou2024}. While recent systems increasingly use iterative reasoning, fine-tuning, retrieval augmentation, structured program representations, and agentic workflows \cite{fakih2025,kim2025,lekssays2025,liu2026securereviewer,liu2025agent,yang2024,xu2025logsage}, it remains unclear which DevSecOps stages receive the most attention, which methodological patterns are emerging, how reported performance should be interpreted, and where important gaps remain.

Several recent reviews have examined LLMs in software engineering, AI-enabled DevSecOps, and LLM-based vulnerability detection and repair \cite{fu2025b,zhou2024,hou2024llmse}. However, these works either cover software engineering broadly, focus on specific security tasks, or examine AI-enabled DevSecOps beyond the recent LLM-focused period. Motivated by this gap, we provide a systematic mapping of LLM-enabled security research across the DevSecOps lifecycle, focusing on studies published or made available between 2024 and 2026 (until June). We analyze the selected studies by lifecycle stage, security task, model family, adaptation strategy, evaluation setting, reported performance, and key limitations. The corresponding research questions are defined in Section~\ref{sec:methods}.

The main contributions of this study are as follows: \begin{itemize} \item We provide a recent lifecycle-oriented mapping of LLM-enabled security research across the six DevSecOps stages. \item We characterize existing approaches in terms of security tasks, models, adaptation strategies, evaluation settings, reported performance, and limitations. \item We synthesize cross-stage patterns in research concentration, model adoption, evaluation maturity, and real-world validation. \item We identify open research directions, including underexplored lifecycle stages, heterogeneous evaluation practices, limited industrial validation, trustworthy autonomous remediation, software supply-chain and Infrastructure-as-Code security, and cross-stage LLM integration. \end{itemize}

The remainder of this paper is organized as follows. Section~\ref{sec:background} presents background and related work. Section~\ref{sec:methods} describes the review methodology and research questions. Section~\ref{sec:landscape_LLM_devsecops} summarizes the overall LLM adoption landscape, and Section~\ref{sec:stage_wise_analysis} provides the stage-wise and cross-stage analysis. 
Section~\ref{sec:future} discusses research gaps and future directions, Section~\ref{sec:limitation} outlines threats to validity, and Section~\ref{sec:conclusion} concludes the paper.

\section{Background and Motivation}
\label{sec:background}


\subsection{DevSecOps Lifecycle and Security Tasks}

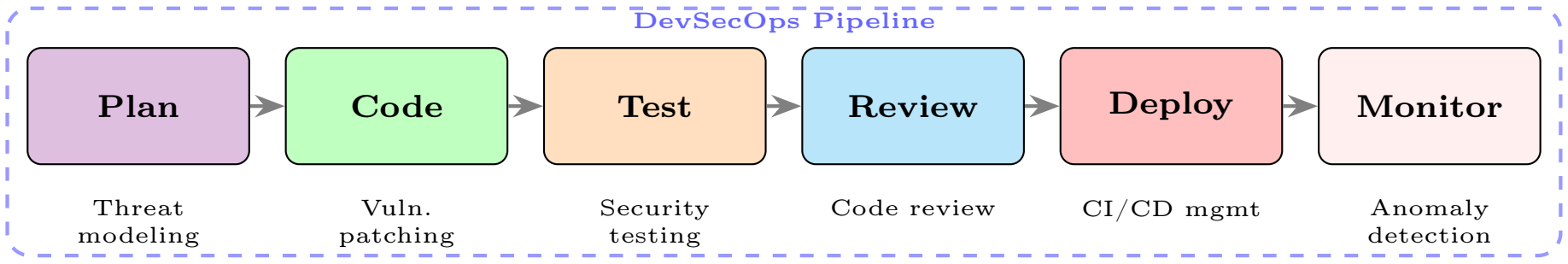
\begin{figure}[h]
\centering
\begin{tikzpicture}[
  stage/.style={rectangle, draw, rounded corners=3pt, minimum width=1.6cm,
                minimum height=0.9cm, align=center, font=\scriptsize\bfseries,
                text width=1.5cm},
  arr/.style={-{Stealth}, thick, gray},
  label/.style={font=\tiny, align=center, text width=1.5cm}
]
\node[stage, fill=violet!25] (plan)    at (0,0)    {Plan};
\node[stage, fill=green!25]  (code)    at (2.0,0)  {Code};
\node[stage, fill=orange!25] (test)    at (4.0,0)  {Test};
\node[stage, fill=cyan!25]   (review)  at (6.0,0)  {Review};
\node[stage, fill=red!25]    (deploy)  at (8.0,0)  {Deploy};
\node[stage, fill=pink!25]   (monitor) at (10.0,0) {Monitor};

\draw[arr] (plan) -- (code);
\draw[arr] (code) -- (test);
\draw[arr] (test) -- (review);
\draw[arr] (review) -- (deploy);
\draw[arr] (deploy) -- (monitor);

\node[label, below=0.15cm of plan]    {\tiny Threat modeling};
\node[label, below=0.15cm of code]    {\tiny Vuln. patching};
\node[label, below=0.15cm of test]    {\tiny Security testing};
\node[label, below=0.15cm of review]  {\tiny Code review};
\node[label, below=0.15cm of deploy]  {\tiny CI/CD mgmt};
\node[label, below=0.15cm of monitor] {\tiny Anomaly detection};

\draw[draw=blue!40, dashed, rounded corners=5pt, thick]
  ($(plan.north west)+(-0.15,0.3)$) rectangle ($(monitor.south east)+(0.15,-0.7)$);
\node[font=\tiny\bfseries, text=blue!60] at (5.0, 0.65) {DevSecOps Pipeline};

\end{tikzpicture}
\caption{The DevSecOps pipeline showing six stages and associated security tasks. Security is embedded at every stage rather than applied as a final gate.}
\label{fig:devsecops}
\end{figure}

DevSecOps integrates security throughout software development and operations rather than treating it as a separate activity performed near the end of the SDLC \cite{fu2025b}. In this work, we organize the DevSecOps lifecycle into six stages 
as illustrated in Fig.~\ref{fig:devsecops}. Each stage involves different security activities. The Plan stage includes security requirements analysis, threat modeling, and attack-surface assessment; Code focuses on secure implementation, vulnerability identification, and repair; Test includes vulnerability detection, security testing, and exploit validation; Review covers security-aware code and configuration review; Deploy includes CI/CD security, deployment validation, and failure diagnosis; and Monitor focuses on runtime anomaly detection, incident analysis, compliance monitoring, and remediation \cite{fu2025b}. Although these stages provide a useful organizational structure, modern DevSecOps pipelines are continuous, and information generated at one stage may affect security decisions at subsequent stages.

\subsection{LLMs and Language-model-based Methods for Software Security}

LLMs have expanded the role of language-based models in software engineering from code completion and generation to more complex security-oriented reasoning and automation \cite{zhang2025llms,hou2024llmse}. Recent studies use LLMs for tasks such as automated threat modeling \cite{bedoya2024,yang2024}, vulnerability detection and repair \cite{fakih2025,kim2025,lekssays2025,nong2025}, security code review \cite{liu2026securereviewer,yu2026}, CI/CD failure diagnosis \cite{xu2025logsage}, and runtime anomaly detection and remediation \cite{guan2025,jha2025,kakarla2024}. These systems also differ substantially in how the underlying models are adapted or integrated into security workflows. Existing approaches include direct and reasoning-oriented prompting \cite{kim2025,yang2024}, iterative workflows \cite{fakih2025,kim2025}, parameter-efficient fine-tuning such as LoRA \cite{liu2026securereviewer,yang2024}, retrieval-augmented generation \cite{xu2025logsage}, structured program representations such as code property graphs \cite{lekssays2025}, and multi-agent or agentic architectures \cite{cui2024,liu2025agent}.

This methodological diversity makes model-level comparison alone insufficient for understanding the current landscape. The effectiveness of an LLM-enabled security system can depend not only on the base model but also on the contextual information provided to the model, the adaptation strategy, the surrounding tools or representations, and the evaluation setting. Moreover, LLM integration introduces additional security and trustworthiness concerns, including hallucination, unreliable reasoning, privacy exposure, prompt- or agent-level attacks, and risks associated with automated actions in security-critical environments \cite{das2025security,gupta2023chatgpt,shibli2024abusegpt}. Therefore, understanding LLM adoption in DevSecOps requires examining both the models themselves and the broader systems in which they operate.

\begin{table}[t]
\centering
\caption{Positioning of this study relative to representative recent reviews.}
\label{tab:related_surveys}
\small
\begin{tabular}{|p{2.2cm}| p{4cm}| p{1.9cm}| c | c|}
\hline
\textbf{Study} & \textbf{Primary Scope} & \textbf{Focus} & \textbf{Lifecycle-} & \textbf{LLM-} \\
 & &  & \textbf{wide} & \textbf{specific}\\
\hline
Hou et al. \cite{hou2024llmse} & LLMs in software engineering & General SE & \checkmark & \checkmark \\
\hline
Fu et al. \cite{fu2025b} & AI in DevSecOps & Security & \checkmark & -- \\
\hline
Zhou et al. \cite{zhou2024} & Vulnerability detection and repair & Security & -- & \checkmark \\
\hline
Zhang et al. \cite{zhang2025llms} & LLMs in cybersecurity & Security & -- & \checkmark \\
\hline
\textbf{This study} & \textbf{LLMs across DevSecOps} & \textbf{Security} & \checkmark & \checkmark \\
\hline
\end{tabular}
\end{table}

\subsection{Existing Reviews and Research Gap}

Recent reviews have examined LLMs in software engineering, cybersecurity, vulnerability detection and repair, and AI-enabled DevSecOps \cite{fu2025b,hou2024llmse,zhang2025llms,zhou2024}. Collectively, these studies provide important foundations, but their scopes differ from the objective of this work. Broad software-engineering reviews include many non-security tasks, security-focused surveys often concentrate on specific problems, and AI-for-DevSecOps reviews cover a wider range of AI techniques beyond the recent LLM-focused landscape.

As summarized in Table~\ref{tab:related_surveys}, this study instead provides a recent, lifecycle-oriented analysis of LLM-enabled security research published or made available from January 2024 through May 2026. We examine how LLMs are applied across all six DevSecOps stages and compare their security tasks, model families, adaptation strategies, evaluation settings, reported effectiveness, and limitations. This perspective enables analysis not only of where research activity is concentrated, but also of differences in evaluation maturity and real-world validation across lifecycle stages.

\section{Review Methodology}
\label{sec:methods}

This review examines recent LLM-enabled security research across the DevSecOps lifecycle, focusing on studies published from 2024 through 2026 (until June). We follow the PRISMA 2020 \cite{prisma2020} review process to identify, screen, and analyze relevant studies. The selected studies are mapped to the six DevSecOps stages: Plan, Code, Test, Review, Deploy, and Monitor, which are characterized in terms of security tasks, models, adaptation strategies, evaluation settings, reported performance, strengths, and limitations.

\subsection{Research Questions and Scope}

This paper addresses four research questions: \textbf{RQ1:} What security tasks are addressed by LLM-enabled research across the six DevSecOps stages during 2024--2026? \textbf{RQ2:} What LLMs, supporting language models, adaptation strategies, and architectural methods are used within each stage? \textbf{RQ3:} How are these approaches evaluated, what performance is reported, and what strengths and limitations emerge? \textbf{RQ4:} What gaps remain in lifecycle coverage, evaluation maturity, industrial validation, trustworthy autonomous remediation, software supply-chain and Infrastructure-as-Code security, and cross-stage LLM integration?

We define an LLM-enabled study as one in which a generative or instruction-capable LLM constitutes a substantive component of the proposed security method or evaluation. Encoder-based models such as BERT or CodeBERT are treated as supporting language models when used within hybrid LLM architectures. Studies relying exclusively on conventional machine learning or encoder-only NLP models are excluded from the primary corpus, although they may be discussed as baselines or supporting technologies.

\subsection{Search and Study Selection}

Relevant studies were searched in IEEE Xplore, ACM Digital Library, Springer, Scopus, USENIX publications, and arXiv. Search queries combined LLM-related terms with security and DevSecOps concepts, including ``LLM DevSecOps,'' ``LLM vulnerability detection,'' ``LLM vulnerability repair,'' ``LLM secure code review,'' ``LLM security testing,'' ``LLM threat modeling,'' ``LLM CI/CD security,'' and ``LLM anomaly detection.''

Studies were included if they: (1) were published between 2024 and 2026; (2) addressed a security task associated with at least one DevSecOps stage; (3) contained a substantive LLM component; (4) provided empirical evaluation, experimental analysis, or a detailed case study; and (5) provided sufficient information to characterize the proposed method and evaluation. Studies outside the review period, unrelated to software or DevSecOps security, based solely on encoder-only NLP or conventional machine learning, or lacking empirical evidence were excluded. Duplicate studies were also removed.

The remaining studies were screened by title and abstract, followed by full-text assessment. Eligible studies were then mapped to their primary DevSecOps stage according to the dominant security task. Studies spanning multiple stages were assigned a primary stage while retaining their cross-stage characteristics for later analysis.

\subsection{Data Extraction and Coding}

For each selected study, we extracted the publication year, DevSecOps stage, security task, primary LLM, supporting models or tools, adaptation strategy, dataset or data source, evaluation setting, evaluation metrics, reported performance, strengths, limitations, and artifact availability. Adaptation strategies include direct or few-shot prompting, structured or iterative prompting, reasoning approaches, retrieval-augmented generation, fine-tuning or LoRA, tool augmentation, and agentic or multi-agent methods.

Evaluation settings were additionally characterized by their level of operational realism, ranging from prototype or benchmark evaluation to realistic repository/system evaluation and industrial or production validation. This distinction enables analysis of not only reported performance but also the maturity of the supporting evidence.

\subsection{Analysis Strategy}

The analysis is conducted at two levels. First, a stage-wise analysis compares the security tasks, methods, models, adaptation strategies, evaluation settings, reported performance, strengths, and weaknesses within each DevSecOps stage. Second, a cross-stage synthesis examines research concentration, model adoption, methodological trends, evaluation heterogeneity, evidence maturity, and industrial validation across the 2024--2026 (until June) corpus.

Because the reviewed studies address different tasks and frequently use different datasets, baselines, and performance metrics, reported results are interpreted within the experimental context of each study rather than used to construct a global ranking of models or systems.

\section{Lifecycle Landscape of LLM-Enabled DevSecOps Research}
\label{sec:landscape_LLM_devsecops}

This section maps the reviewed 2024--2026 (until June 2026) literature to the six DevSecOps stages and summarizes the security tasks receiving recent research attention. The lifecycle-oriented mapping provides the basis for the detailed analysis of methods, models, performance, strengths, and limitations presented in Section~5.

\subsection{Mapping Across the DevSecOps Lifecycle}

We organize the selected studies according to the primary security task addressed within the DevSecOps lifecycle. The \textit{Plan} stage covers threat modeling, attack modeling, and security planning \cite{bedoya2024,marin2026,yang2024}. The \textit{Code} stage includes vulnerability repair, patch generation, and security-patch analysis \cite{fakih2025,kim2025,li2026,nong2025}. The \textit{Test} stage covers vulnerability detection, exploit validation, security testing, fuzzing, and protocol verification \cite{abdiukov2024,lekssays2025,liu2025agent,lotfi2025,song2026}. The \textit{Review} stage focuses primarily on security-aware code review and analysis of development or CI/CD artifacts \cite{chomatek2025,liu2026securereviewer,yu2026}. The \textit{Deploy} stage includes CI/CD diagnosis, pipeline optimization, and deployment remediation \cite{cui2024,xu2025logsage}, while the \textit{Monitor} stage addresses runtime anomaly detection, log analysis, compliance monitoring, and incident response \cite{guan2025,jha2025,kakarla2024,kolagani2025}. Table~\ref{tab:stage_distribution} summarizes the resulting lifecycle mapping.

\begin{table}[t]
\centering
\caption{Distribution of reviewed studies ($N_{total}$= 21) across DevSecOps stages.}
\label{tab:stage_distribution}
\small
\begin{tabular}{|p{1.2cm} | p{6.5cm} | c|}
\hline
\textbf{Stage} & \textbf{Representative Security Tasks} & \textbf{N} \\
\hline

Plan & Threat modeling, attack modeling, security planning & 3 \\
\hline
Code & Vulnerability repair, patch generation, patch analysis & 4 \\
\hline
Test & Vulnerability detection, exploit validation, security testing, protocol verification & 5 \\
\hline
Review & Security-aware code and pipeline review & 3 \\
Deploy & CI/CD diagnosis, optimization, and remediation & 2 \\
\hline
Monitor & Log analysis, anomaly detection, compliance, incident response & 4 \\
\hline
\end{tabular}
\end{table}

\subsection{Research Distribution and Task Coverage}

The primary LLM-enabled literature is unevenly distributed across the six DevSecOps stages. The \textit{Test} stage receives the highest coverage with five studies, followed closely by \textit{Code} and \textit{Monitor} with four studies each. \textit{Plan} and \textit{Review} are represented by three studies each, while \textit{Deploy} remains the least studied stage with two studies. Thus, recent LLM-enabled security research is concentrated primarily in testing, secure coding, and operational monitoring, although no single lifecycle stage dominates the current literature.

Task coverage also varies across stages. Test-stage studies address vulnerability detection, exploit validation, security testing, agent security, and protocol verification, while Code-stage studies focus mainly on vulnerability repair, patch generation, and security-patch analysis. Monitor-stage research emphasizes log analysis, anomaly detection, compliance monitoring, incident response, and remediation. Plan-stage work primarily addresses threat and adversary modeling, whereas Review focuses on security-aware code and pipeline analysis. Deploy-stage studies remain comparatively limited and concentrate on CI/CD diagnosis, optimization, and remediation.

Overall, the distribution indicates relatively stronger recent activity in \textit{Test, Code, and Monitor}, moderate coverage of \textit{Plan and Review}, and limited work in \textit{Deploy}. In addition, areas such as software supply-chain security, Infrastructure-as-Code analysis, and cross-stage LLM orchestration remain sparsely represented or absent from the primary corpus. Benchmark-oriented studies that do not satisfy the primary LLM inclusion criterion are considered separately when analyzing evaluation quality and methodological limitations.

\section{Stage-Wise and Cross-Stage Analysis of LLM-Enabled DevSecOps}
\label{sec:stage_wise_analysis}

This section analyzes the 21 primary LLM-enabled studies across the six DevSecOps stages, focusing on the main security tasks, models, adaptation strategies, evaluation evidence, and stage-level findings. Because the reviewed studies use heterogeneous datasets, baselines, and metrics, reported performance is interpreted within each study's experimental context rather than used for direct global ranking.

\subsection{Plan: Threat Modeling and Security Planning}

The three Plan-stage studies focus mainly on automated threat modeling and adversary analysis \cite{bedoya2024,marin2026,yang2024}. Existing approaches use GPT-family, Llama, and DeepSeek models with structured prompting, optimized reasoning, LoRA-based adaptation, and automated attack-defense tree generation. Bedoya et al. \cite{bedoya2024} provide evidence from a realistic cloud-based organizational system, while Threat Modeling-LLM \cite{yang2024} shows that reasoning-oriented prompting and task-specific adaptation can improve STRIDE-based threat modeling over direct prompting. Mar\'{i}n-Caballero et al. \cite{marin2026} further extend LLM-assisted adversary modeling to cloud-native environments.

The main takeaway is that Plan-stage research is moving from generic prompting toward \textit{structured and domain-adapted security reasoning}. However, evidence remains concentrated in specific application domains and relatively limited evaluation settings, leaving broader generalizability insufficiently established.

\subsection{Code: Vulnerability Repair and Secure Coding}

The four Code-stage studies primarily address vulnerability repair, patch generation, and security-patch analysis \cite{fakih2025,kim2025,li2026,nong2025}. A consistent methodological pattern is the use of iterative, adaptive, or reasoning-guided workflows rather than one-shot generation. LLM4CVE \cite{fakih2025} uses iterative feedback for vulnerability repair, SAN2PATCH \cite{kim2025} applies Tree-of-Thought reasoning and reports a 79.5\% patch success rate in its evaluation setting, and AppPatch \cite{nong2025} uses adaptive prompting for previously unseen vulnerabilities. Li et al. \cite{li2026} extend LLM use beyond patch generation toward security-patch categorization and prioritization.

The stage-level evidence suggests that \textit{multi-step reasoning and feedback improve automated repair effectiveness}. At the same time, evaluation remains heavily dependent on benchmarks, open-source repositories, specific programming languages, or single software ecosystems, limiting claims about production-scale generalization.

\subsection{Test: Security Testing and Vulnerability Validation}

The Test stage contains five primary studies and exhibits the greatest methodological diversity \cite{abdiukov2024,lekssays2025,liu2025agent,lotfi2025,song2026}. Rather than relying on LLMs alone, several approaches combine language-model reasoning with program-analysis and testing mechanisms. LLMxCPG \cite{lekssays2025} integrates LLM reasoning with Code Property Graphs and reports substantial reduction of irrelevant input, while ProtocolGuard \cite{song2026} combines LLM-guided static analysis with dynamic verification. AgentFuzz \cite{liu2025agent} extends testing to vulnerabilities introduced by LLM-based agents themselves, and other studies investigate exploit validation and integration with established security-testing techniques \cite{abdiukov2024,lotfi2025}.

A key finding is that Test-stage research is evolving toward \textit{hybrid LLM--security analysis architectures}, where structured program context and conventional analysis tools complement LLM capabilities. At the same time, benchmark quality remains an important concern. PrimeVul \cite{ding2024}, although not counted in the primary corpus, identifies duplication, labeling errors, and data leakage that can distort vulnerability-detection results.

\subsection{Review: Security-Aware Code and Pipeline Review}

The three Review-stage studies examine security-aware review of code and development artifacts \cite{chomatek2025,liu2026securereviewer,yu2026}. The literature emphasizes two main directions: reasoning-oriented model selection and domain-specific adaptation. Yu et al. \cite{yu2026} compare multiple LLMs for security code review and report stronger performance from reasoning-oriented models, while SecureReviewer \cite{liu2026securereviewer} shows that security-aware fine-tuning and LoRA can outperform zero-shot prompting. Chom\k{a}tek et al. \cite{chomatek2025} provide complementary large-scale evidence on security practices across more than 28,000 CI/CD configurations using LLM-based analysis.

The main takeaway is that \textit{reasoning capability and security-specific adaptation appear important for effective code review}. However, the stage remains relatively small and relies primarily on open-source or non-production evaluations, making industrial readiness difficult to assess.

\subsection{Deploy: CI/CD Security and Deployment Remediation}

Deploy is the least represented stage, with two studies focused on CI/CD diagnosis, optimization, and remediation \cite{cui2024,xu2025logsage}. Despite the small research base, it contains some of the strongest operational evidence. LogSage \cite{xu2025logsage} combines structured prompting and retrieval-augmented generation and is evaluated on approximately 1.07 million CI/CD executions at ByteDance, reporting more than 98\% root-cause-analysis precision. Cui \cite{cui2024} explores a multi-agent architecture for autonomous CI/CD optimization.

The emerging pattern is a transition toward \textit{retrieval-grounded diagnosis and agentic operational automation}. The limited number of studies, however, makes it difficult to determine whether these results generalize across organizations, infrastructure stacks, and CI/CD platforms.

\subsection{Monitor: Runtime Monitoring and Incident Response}

The Monitor stage contains four studies, equal in coverage to Code, and focuses on log analysis, anomaly detection, compliance monitoring, incident response, and remediation \cite{guan2025,jha2025,kakarla2024,kolagani2025}. Approaches range from hybrid LLM architectures to increasingly autonomous remediation. LogLLM \cite{guan2025}, for example, combines BERT-based representations with Llama for log anomaly detection, while Kakarla \cite{kakarla2024} proposes an autonomous remediation framework with human oversight. Jha et al. \cite{jha2025} provide comparatively strong operational evidence through deployment with more than 500 users and over 200,000 API calls.

Unlike Code, which focuses largely on software artifacts before deployment, Monitor-stage research emphasizes \textit{continuous operational decision support and increasingly autonomous response}. This also introduces additional concerns regarding governance, auditability, rollback, and human oversight when LLMs are allowed to influence or execute remediation actions. Table~\ref{tab:stage_summary} summarizes the dominant methodological patterns and limitations across the six DevSecOps stages.


\begin{table*}[t]
\centering
\caption{Summary of dominant LLM-enabled methods and limitations across DevSecOps stages.}
\label{tab:stage_summary}
\small
\begin{tabular}{|p{1.1cm}|p{3.0cm}|p{4.1cm}|p{4.0cm}|}
\hline
\textbf{Stage} & \textbf{Main Tasks} & \textbf{Dominant Methods} & \textbf{Main Limitations} \\
\hline
Plan & Threat and adversary modeling & Structured prompting, reasoning-oriented approaches, and LoRA-based adaptation & Domain-specific evaluations and limited evidence of broader generalization \\
\hline
Code & Vulnerability repair, patch generation, and patch analysis & Iterative prompting, adaptive prompting, and reasoning-guided generation & Benchmark-heavy evaluation, programming-language dependence, and limited production validation \\
\hline
Test & Vulnerability detection, validation, and security testing & Hybrid LLM--program analysis, structured program representations, and static/dynamic verification & Benchmark-quality concerns, architectural complexity, and limited operational validation \\
\hline
Review & Security-aware code and pipeline review & Multi-model reasoning and security-specific fine-tuning & Small research base, open-source-heavy evaluation, and limited industrial validation \\
\hline
Deploy & CI/CD diagnosis, optimization, and remediation & Retrieval-augmented generation, structured diagnostic prompting, and multi-agent systems & Few studies, proprietary environments, and limited cross-organizational evidence \\
\hline
Monitor & Log analysis, anomaly detection, incident response, and remediation & Hybrid language-model architectures, LLM reasoning, and autonomous remediation & Reproducibility limitations and unresolved governance and autonomous-action risks \\
\hline
\end{tabular}
\end{table*}

\subsection{Cross-Stage Findings and Emerging Trends}

Three broader findings emerge across the 2024--2026 literature. First, the field is moving beyond standalone or direct LLM prompting toward \textit{structured, task-adapted, and hybrid approaches}. Iterative repair, reasoning-guided prompting, LoRA-based adaptation, hybrid program analysis, retrieval-augmented generation, and agentic architectures appear across multiple stages \cite{fakih2025,kim2025,lekssays2025,liu2026securereviewer,nong2025,yang2024,xu2025logsage}. This suggests that effectiveness increasingly depends on how LLMs are integrated with security-specific context and existing tools rather than on model choice alone.

Second, \textit{evaluation maturity remains uneven}. Most studies rely on benchmarks, open-source repositories, or controlled environments, whereas only a small number provide substantial operational evidence. LogSage \cite{xu2025logsage}, Jha et al. \cite{jha2025}, and Bedoya et al. \cite{bedoya2024} demonstrate that real-world deployment is feasible, but such evaluations remain uncommon.

Third, \textit{research volume does not directly correspond to evidence maturity}. Test has the largest number of primary studies, while Code and Monitor have comparable coverage, yet some of the strongest operational evidence appears in Deploy and Monitor. Overall, the literature shows a transition toward more specialized, integrated, and autonomous LLM-enabled DevSecOps systems, while heterogeneous evaluation and limited industrial validation remain major barriers to assessing generalizability and deployment readiness.


\section{Open Research Directions}
\label{sec:future}


The 2024--2026 literature shows rapid progress in LLM-enabled DevSecOps, but several gaps remain before these systems can be considered broadly reliable and deployable.

\noindent\textbf{Underexplored Lifecycle Stages and Security Tasks.} Research coverage remains uneven across the DevSecOps lifecycle. Deploy is the least studied stage, while Plan and Review also receive less attention than Test, Code, and Monitor. Future work should expand beyond vulnerability detection and repair toward deployment-time security, security-aware review, and operational assurance. Software supply-chain analysis, dependency and provenance checking, secrets management, and Infrastructure-as-Code security also remain largely absent despite their importance to modern CI/CD and cloud-native environments.

\noindent\textbf{Standardized and Realistic Evaluation.} Heterogeneous datasets, metrics, baselines, and evaluation environments make cross-study comparison difficult. Future work should adopt shared benchmarks and reporting practices that capture not only task-specific effectiveness but also robustness, latency, computational cost, and failure behavior. Benchmark quality also requires attention because duplication, labeling errors, and data leakage can distort reported results \cite{ding2024}. Evaluation should increasingly move beyond offline benchmarks and open-source repositories toward realistic development settings, production workloads, and longitudinal deployments. The limited industrial evidence in the current corpus \cite{bedoya2024,jha2025,xu2025logsage} suggests that deployment maturity still lags methodological progress.

\noindent\textbf{Trustworthy Autonomous Remediation.} Recent work increasingly moves from passive assistance toward automated or agentic remediation \cite{cui2024,kakarla2024,xu2025logsage}. Because incorrect actions may directly modify code, pipelines, configurations, or production systems, future research should address authorization, human approval, confidence estimation, auditability, rollback, and runtime verification. Trustworthiness must also include robustness to hallucination, malicious context, prompt injection, and unsafe tool use, making protection of the LLM-enabled workflow itself a critical requirement.

\noindent\textbf{Cross-Stage LLM Integration.} Most current approaches remain stage-specific, leaving security knowledge fragmented across the lifecycle. Future systems should preserve and reuse security context across Plan, Code, Test, Review, Deploy, and Monitor. For example, a runtime anomaly could trigger root-cause analysis, guide a code repair, initiate validation and review, and support controlled redeployment. Such integration could enable continuous security feedback, but raises challenges in context propagation, authorization boundaries, error accumulation, and coordination among multiple LLMs or agents.

\noindent\textbf{Human Oversight, Explainability, and Operational Trust.} As LLMs increasingly influence security decisions, practitioners need to understand why vulnerabilities are flagged, patches are recommended, or remediation actions are executed. Explainability is therefore important for trust and accountability, particularly in high-impact or regulated environments. Existing work provides limited evidence on how explanations affect human decision-making or whether model reasoning reliably corresponds to correct security behavior \cite{cambria2024xai,Maraz_TPS_2025_explainable}. Future studies should examine when human approval is required, how uncertainty should be communicated, and how operators can verify or override model-generated decisions.

\noindent\textbf{Toward Production-Grade LLM-Enabled DevSecOps.} Overall, future research should move beyond isolated LLM capabilities toward systems that can be evaluated, governed, and integrated reliably across real DevSecOps environments.

\section{Threats to Validity and Review Limitations}
\label{sec:limitation}

This review focuses on studies published during 2024--2026 (until June) to capture recent LLM-enabled DevSecOps trends; therefore, earlier influential work may be excluded. The search is also limited by the selected databases, keywords, and terminology used in the literature. Some classification subjectivity remains because studies may span multiple DevSecOps stages; we assign each study based on its dominant security task. To maintain scope consistency, encoder-only NLP studies are excluded from the primary corpus unless they are part of a broader architecture with a substantive generative or instruction-capable LLM component. Finally, heterogeneous tasks, datasets, metrics, and baselines limit direct performance comparison, while the small number of industrial evaluations constrains conclusions about real-world generalizability. The findings should therefore be interpreted as a snapshot of the 2024--2026 research landscape.

\section{Conclusion}
\label{sec:conclusion}


This review maps recent LLM-enabled security research across the six DevSecOps stages and shows that activity is concentrated in Test, Code, and Monitor, while Deploy remains comparatively underexplored. Across stages, research is shifting from direct prompting toward structured reasoning, iterative and adaptive methods, domain adaptation, hybrid analysis, retrieval augmentation, and agentic architectures.

At the same time, no single model or method dominates across the lifecycle, and evaluation remains highly heterogeneous. Most studies rely on benchmarks, open-source repositories, or controlled environments, with relatively limited industrial validation. Future progress will therefore depend on more standardized and realistic evaluation, trustworthy autonomous remediation, stronger coverage of software supply-chain and Infrastructure-as-Code security, improved human oversight, and cross-stage integration of LLM-enabled security capabilities.


\bibliographystyle{splncs04}
\bibliography{reference}

\begin{thebibliography}{10}
\providecommand{\url}[1]{\texttt{#1}}
\providecommand{\urlprefix}{URL }
\providecommand{\doi}[1]{https://doi.org/#1}

\bibitem{abdiukov2024}
Abdiukov, T.: Automated security testing in {DevSecOps} pipelines: Integrating {AI}-based vulnerability discovery and compliance validation. World Journal of Advanced Research and Reviews  \textbf{22}(01),  2083--2093 (2024). \doi{10.30574/wjarr.2024.22.1.1083}, \url{https://doi.org/10.30574/wjarr.2024.22.1.1083}

\bibitem{bedoya2024}
Bedoya, M., Palacios, S., D{\'\i}az-L{\'o}pez, D., Laverde, E., Nespoli, P.: Enhancing {DevSecOps} practice with large language models and security chaos engineering. International Journal of Information Security  \textbf{23},  3765--3788 (2024). \doi{10.1007/s10207-024-00909-w}, \url{https://doi.org/10.1007/s10207-024-00909-w}

\bibitem{cambria2024xai}
Cambria, E., Malandri, L., Mercorio, F., Nobani, N., Seveso, A.: Xai meets llms: A survey of the relation between explainable ai and large language models. arXiv preprint arXiv:2407.15248  (2024)

\bibitem{chomatek2025}
Chom\k{a}tek, L., Papuga, J., Nowak, P., Poniszewska-Mara\'{n}da, A.: Decoding ci/cd practices in open-source projects with llm insights. In: Proceedings of the 33rd ACM International Conference on the Foundations of Software Engineering. pp. 1638--1644. FSE Companion '25, Association for Computing Machinery, New York, NY, USA (2025). \doi{10.1145/3696630.3728699}, \url{https://doi.org/10.1145/3696630.3728699}

\bibitem{cui2024}
Cui, J.: Generative {AI} in {DevOps}: Autonomous {CI/CD} pipeline optimization via {LLM}-based multi-agent systems. In: Proceedings of the International Conference on Innovative Data Communication Technologies and Application. Solbridge International School of Business, Woosong University, Daejeon, Republic of Korea (2024), \url{https://www.researchgate.net/publication/393965880}

\bibitem{das2025security}
Das, B.C., Amini, M.H., Wu, Y.: Security and privacy challenges of large language models: A survey. ACM Computing Surveys  \textbf{57}(6),  1--39 (2025)

\bibitem{ding2024}
Ding, Y., Fu, Y., Ibrahim, O., Sitawarin, C., Chen, X., Alomair, B., Wagner, D., Ray, B., Chen, Y.: Vulnerability detection with code language models: How far are we? arXiv preprint arXiv:2403.18624  (2024), \url{https://arxiv.org/abs/2403.18624}

\bibitem{fakih2025}
Fakih, M., Dharmaji, R., Bouzidi, H., Quiros~Araya, G., Ogundare, O., Al~Faruque, M.A.: {LLM4CVE}: Enabling iterative automated vulnerability repair with large language models. arXiv preprint arXiv:2501.03446  (2025), \url{https://arxiv.org/abs/2501.03446}

\bibitem{fu2025b}
Fu, M., Pasuksmit, J., Tantithamthavorn, C.: Ai for devsecops: A landscape and future opportunities. ACM Trans. Softw. Eng. Methodol.  \textbf{34}(4) (2025). \doi{10.1145/3712190}, \url{https://doi.org/10.1145/3712190}

\bibitem{guan2025}
Guan, W., Cao, J., Qian, S., Gao, J., Ouyang, C.: {LogLLM}: Log-based anomaly detection using large language models. arXiv preprint arXiv:2411.08561  (2025), \url{https://arxiv.org/abs/2411.08561}

\bibitem{gupta2023chatgpt}
Gupta, M., Akiri, C., Aryal, K., Parker, E., Praharaj, L.: From chatgpt to threatgpt: Impact of generative ai in cybersecurity and privacy. IEEE access  \textbf{11},  80218--80245 (2023)

\bibitem{hou2024llmse}
Hou, X., Zhao, Y., Liu, Y., Yang, Z., Wang, K., Li, L., Luo, X., Lo, D., Grundy, J., Wang, H., et~al.: Large language models for software engineering: A systematic literature review. ACM Transactions on Software Engineering and Methodology  \textbf{33}(8),  1--79 (2024). \doi{10.1145/3695988}

\bibitem{jha2025}
Jha, N., Lin, S., Jayaraman, S., Frohling, K., Constantinides, C., Patel, D.: {LLM} assisted anomaly detection service for site reliability engineers: Enhancing cloud infrastructure resilience. arXiv preprint arXiv:2501.16744  (2025), \url{https://arxiv.org/abs/2501.16744}

\bibitem{kakarla2024}
Kakarla, R.: {LLM}-based autonomous remediation for {DevSecOps} pipelines. Eastasouth Journal of Information System and Computer Science  \textbf{2}(2),  179--188 (2024). \doi{10.58812/esiscs.v2i02}, \url{https://doi.org/10.58812/esiscs.v2i02}

\bibitem{kim2025}
Kim, Y., Shin, S., Kim, H., Yoon, J.: Logs in, patches out: Automated vulnerability repair via tree-of-thought {LLM} analysis. In: Proceedings of the 34th USENIX Security Symposium (USENIX Security 25). pp. 4401--4419. USENIX Association, Seattle, WA, USA (2025), \url{https://www.usenix.org/conference/usenixsecurity25/presentation/kim-youngjoon}

\bibitem{kolagani2025}
Kolagani, S.H.D., Bhandar, M., Altounjy, R.: Enhancing {DevOps} security with {LLMs} for automation and compliance. In: Proceedings of the 7th International Conference on Innovative Data Communication Technologies and Application (ICIDCA-2025). IEEE (2025). \doi{10.1109/ICIDCA66325.2025.11280531}, \url{https://doi.org/10.1109/ICIDCA66325.2025.11280531}

\bibitem{lekssays2025}
Lekssays, A., Mouhcine, H., Tran, K., Yu, T., Khalil, I.: {LLMxCPG}: Context-aware vulnerability detection through code property graph-guided large language models. In: Proceedings of the 34th USENIX Security Symposium (USENIX Security 25). USENIX Association, Seattle, WA, USA (2025), \url{https://www.usenix.org/conference/usenixsecurity25/presentation/lekssays}

\bibitem{li2026}
Li, X., Pu, J., Wu, Y., Zou, X., Zhu, S., Wu, Q., Zhang, Z., Hsu, J., Dong, Y., Qian, Z., Lu, K., Jaeger, T., De~Lucia, M., Krishnamurthy, S.V.: What do they fix? {LLM}-aided categorization of security patches for critical memory bugs. arXiv preprint arXiv:2509.22796  (2025), \url{https://arxiv.org/abs/2509.22796}

\bibitem{liu2026securereviewer}
Liu, F., Liu, S., Zhu, Y., Lian, X., Zhang, L.: {SecureReviewer}: Enhancing large language models for secure code review through secure-aware fine-tuning. In: Proceedings of the 2026 IEEE/ACM 48th International Conference on Software Engineering (ICSE '26). ACM, New York, NY, USA (2026). \doi{10.1145/3744916.3773191}, \url{https://doi.org/10.1145/3744916.3773191}

\bibitem{liu2025agent}
Liu, F., Zhang, Y., Luo, J., Dai, J., Chen, T., Yuan, L., Yu, Z., Shi, Y., Li, K., Zhou, C., Chen, H., Yang, M.: Make agent defeat agent: Automatic detection of taint-style vulnerabilities in {LLM}-based agents. In: Proceedings of the 34th USENIX Security Symposium (USENIX Security 25). USENIX Association, Seattle, WA, USA (2025), \url{https://www.usenix.org/conference/usenixsecurity25/presentation/liu-fengyu}

\bibitem{lotfi2025}
Lotfi, A., Katsis, C., Bertino, E.: Automated vulnerability validation and verification: A large language model approach. arXiv preprint arXiv:2509.24037  (2025), \url{https://arxiv.org/abs/2509.24037}

\bibitem{marin2026}
Mar{\'i}n-Caballero, M., Betancourt, M., D{\'i}az-L{\'o}pez, D., Perales~G{\'o}mez, A.L., Nespoli, P., Mart{\'i}nez~P{\'e}rez, G.: Anticipating adversary behavior in {DevSecOps} scenarios through large language models. arXiv preprint arXiv:2602.14106  (2026), \url{https://arxiv.org/abs/2602.14106}

\bibitem{Maraz_TPS_2025_explainable}
Mia, M., Pritom, M.M.A.: Explainable but vulnerable: Adversarial attacks on xai explanation in cybersecurity applications. In: 2025 IEEE 7th International Conference on Trust, Privacy and Security in Intelligent Systems, and Applications (TPS-ISA). pp. 101--110 (2025). \doi{10.1109/TPS-ISA67132.2025.00020}

\bibitem{nong2025}
Nong, Y., Yang, H., Cheng, L., Hu, H., Cai, H.: {AppPatch}: Automated adaptive prompting large language models for real-world software vulnerability patching. In: Proceedings of the 34th USENIX Security Symposium (USENIX Security 25). pp. 4481--4500. USENIX Association, Seattle, WA, USA (2025), \url{https://www.usenix.org/conference/usenixsecurity25/presentation/nong}

\bibitem{prisma2020}
Page, M.J., McKenzie, J.E., Bossuyt, P.M., Boutron, I., Hoffmann, T.C., Mulrow, C.D., Shamseer, L., Tetzlaff, J.M., Akl, E.A., Brennan, S.E., et~al.: The prisma 2020 statement: An updated guideline for reporting systematic reviews. BMJ  \textbf{372}, ~n71 (2021). \doi{10.1136/bmj.n71}

\bibitem{shibli2024abusegpt}
Shibli, A.M., Pritom, M.M.A., Gupta, M.: Abusegpt: Abuse of generative ai chatbots to create smishing campaigns. In: 2024 12th International Symposium on Digital Forensics and Security (ISDFS). pp.~1--6. IEEE (2024)

\bibitem{song2026}
Song, X., Pei, L., Wu, J., Zeng, Y., He, G., Zuo, C., Liu, X., Zhao, Q., Guo, S.: {ProtocolGuard}: Detecting protocol non-compliance bugs via {LLM}-guided static analysis and dynamic verification. In: Proceedings of the Network and Distributed System Security (NDSS) Symposium 2026. Internet Society, San Diego, CA, USA (2026), \url{https://www.ndss-symposium.org/wp-content/uploads/2026-f521-paper.pdf}

\bibitem{yang2024}
Wu, T., Yang, S., Liu, S., Nguyen, D., Jang, S., Abuadbba, A.: {ThreatModeling-LLM}: Automating threat modeling using large language models for banking system. arXiv preprint arXiv:2411.17058  (2024), \url{https://arxiv.org/abs/2411.17058}

\bibitem{xu2025logsage}
Xu, W., Luo, J., Huang, T., Sui, K., Geng, J., Ma, Q., Akasaka, I., Shi, X., Tang, J., Cai, P.: {LogSage}: An {LLM}-based framework for {CI/CD} failure detection and remediation with industrial validation. arXiv preprint arXiv:2506.03691  (2025), \url{https://arxiv.org/abs/2506.03691}

\bibitem{yu2026}
Yu, J., Liang, P., Fu, Y., Tahir, A., Shahin, M., Wang, C., Cai, Y.: An insight into security code review with {LLMs}: Capabilities, obstacles, and influential factors. arXiv preprint arXiv:2401.16310  (2026), \url{https://arxiv.org/abs/2401.16310}

\bibitem{zhang2025llms}
Zhang, J., Bu, H., Wen, H., Liu, Y., Fei, H., Xi, R., Li, L., Yang, Y., Zhu, H., Meng, D.: When llms meet cybersecurity: A systematic literature review. Cybersecurity  \textbf{8}(1), ~55 (2025)

\bibitem{zhou2024}
Zhou, X., Cao, S., Sun, X., Lo, D.: Large language model for vulnerability detection and repair: Literature review and the road ahead. arXiv preprint arXiv:2404.02525  (2024), \url{https://arxiv.org/abs/2404.02525}

\end{thebibliography}

\end{document}